\documentclass{ws-ijbc}

\usepackage{amsmath,amssymb}
\usepackage{graphicx,subfigure}
\usepackage{bm}
\usepackage{color}
\definecolor{DarkRed}{rgb}{0.45,0.1,0.1}

\begin{document}
\catchline{}{}{}{}{} 

\title{Nonlinear forecasting of the generalised Kuramoto-Sivashinsky equation}

\markboth{H. Gotoda, M. Pradas, \& S. Kalliadasis}{Nonlinear forecasting
of the generalised Kuramoto-Sivashinsky equation}

\author{Hiroshi Gotoda$^{1,2}$, Marc Pradas$^{1}$\footnote{Currently at Department of Mathematics and Statistics,
The Open University, Milton Keynes MK7 6AA, United Kingdom.}, and Serafim Kalliadasis$^{1,\dag}$}

\address{$^{1}$Department of Chemical Engineering, Imperial College London\\
London SW7 2AZ, United Kingdom\\
$^{2}$Department of Mechanical Engineering, Ritsumeikan University\\
Kusatsu, Shiga 525-8577, Japan\\
$^{\dag}$s.kalliadasis@imperial.ac.uk}

\maketitle

\begin{history}
\received{(to be inserted by publisher)}
\end{history}

\begin{abstract}
We study the emergence of pattern formation and chaotic dynamics in the
one-dimensional (1D) generalized Kuramoto-Sivashinsky (gKS) equation by means
of a time-series analysis, in particular a nonlinear forecasting method which
is based on concepts from chaos theory and appropriate statistical methods.
We analyze two types of temporal signals, a local one and a global one,
finding in both cases that the dynamical state of the gKS solution undergoes
a transition from high-dimensional chaos to periodic pulsed oscillations
through low-dimensional deterministic chaos with increasing the control
parameter of the system. Our results demonstrate that the proposed nonlinear
forecasting methodology allows to elucidate the dynamics of the system in
terms of its predictability properties.
\end{abstract}


\section{Introduction}\label{Sec:Intro}

The emergence of complex spatiotemporal behavior in spatially extended systems
(SES) as a result of several mechanisms nonlinearly interacting with each other
has attracted a lot of attention over the last few decades (see
e.g.~\cite{Cross1993,Knobloch2008,Knobloch2010}).
A well-known example of SES is the Kuramoto-Sivashinsky (KS) equation:
\begin{equation}\label{Eq:KS}
\frac{\partial u}{\partial t}+u\frac{\partial u}{\partial x}+\frac{\partial
^{2}u}{\partial x^{2}}+\nu\frac{\partial ^{4}u}{\partial x^{4}}  =  0
\end{equation}%
where $\nu$ is a viscosity damping parameter. This equation is one of the
simplest prototypes exhibiting spatiotemporal chaos that retains some of the
basic ingredients of any nonlinear SES, namely instability/energy production
($\partial^{2}u/\partial x^{2}$), stability/energy dissipation ($\partial
^{4}u/\partial x^{4}$), and nonlinearity  ($u\partial u/\partial x$) that
transfers energy from low to high wave numbers.

Equation (\ref{Eq:KS}) was first derived in the contexts of phase turbulence
in reaction-diffusion systems~\cite{Kuramoto1976}, wrinkled flame front
propagation~\cite{Sivashinsky1977} and unstable drift waves driven by
electron collision in a tokamak~\cite{LaQuey1975}. Subsequently, it has been
shown to be applicable in a wide spectrum of physical settings,
including hydrodynamic (e.g. thin-film)
instabilities~\cite{Sivashinsky1980,Babchin1983,Hooper1985} and optics such
as bright spots formed by self-forcing of the beam profile~\cite{Munkel1996}.
For thin film flows in particular, such as falling films, the KS equation is
obtained via a weakly nonlinear expansion of the 1D Navier-Stokes equations
subject to wall and free-surface boundary conditions and assuming strong
surface tension effects--long waves~\cite{Homsy1974} (derived before
Kuramoto's paper). Not surprisingly, understanding the dynamical behavior of
the KS solution has been a major topic of research. For example, it has been
shown that decreasing the viscosity damping coefficient $\nu$  in small
system sizes with periodic boundary conditions leads to a transition from
steady states to chaos through a period-doubling bifurcation process, which
is known as Feigenbaum scenario~\cite{Smyrlis1991}.

With the addition of dispersion, the KS equation becomes the gKS equation:
\begin{equation}\label{Eq:gKS}
\frac{\partial u}{\partial t}+u\frac{\partial u}{\partial x}+\frac{\partial
^{2}u}{\partial x^{2}}+\delta \frac{\partial ^{3}u}{\partial x^{3}}+
\nu\frac{\partial ^{4}u}{\partial x^{4}}  =  0
\end{equation}%
where $\delta$ is a positive parameter that characterizes the relative
importance of dispersion. This equation then embraces all elements of any
nonlinear process that involves wave evolution in 1D, and has been reported
in a wide variety of contexts, including a reactive falling
film~\cite{Phil2004}, a film falling down a uniformly heated wall
\cite{Kalliadasis2003} and a film falling down a vertical
fiber~\cite{Christian_PRE} (see also the recent
monograph~\cite{Kalliadasis2012} for the derivation of the KS and gKS
equations for thin-film flows), in step dynamics~\cite{Sato1995,Misbah1996,Sato1998}.
An extended version that includes a non-local term has also been derived for
conducting liquid film in the presence of an external electric field~\cite{Tseluiko2010}.

It is now well-known that for small values of $\delta$ the dynamics of the
gKS equation resembles the spatiotemporal chaotic behavior of the KS
solution, while sufficiently large values of the dispersion parameter tend to
arrest it in favor of spatially periodic travelling
waves~\cite{Kawahara1983}. In a regime of moderate values of $\delta$
however, traveling waves or pulses appear to be randomly interacting with
each other giving rise to what is widely known as weak/dissipative turbulence
(in the ``Manneville sense"~\cite{Manneville1985,Kalliadasis2012}). The
dynamics and interaction of such pulses have been studied extensively via
e.g.~coherent-structure theories~\cite{Balmforth1995,Chang1995,Ei1994,Duprat2009,Tse_PRO,Tse2010a,Duprat2011,Tseluiko19092012} which have been extended to more involved equations~\cite{Marc_PoF,Marc_IMA}
while the regularizing effects of dispersion have been considered by Chang
\textit{et al}.~\cite{Chang1993}, who constructed bifurcation diagrams for
the periodic and solitary solutions of the gKS equation.

Hence, the gKS equation constitutes a paradigmatic model for the
transition from chaotic to regular behaviour in spatially extended systems
and as such it has been extensively analysed from a spatiotemporal analysis
point of view (such transitions are nowadays relatively well-known) but it
has not received attention from the point of view of pure temporal signal
analysis. In fact, there is not up to date a detailed analysis of the chaotic properties of the
gKS solution based on time-series tools (chaos theory), in particular
nonlinear forecasting methodologies, and how dispersion affects e.g.~its
predictability properties (there are, on the other hand, several studies on
the chaotic behavior of both the KS
equation~\cite[e.g.][]{Chian2002,Rempel2003} and a damped version of
it~\cite[see e.g.][]{Elder1997,Rempel2007}). We note that temporal signals
are very often the only experimentally accessible data, hence the importance
of understanding their dynamics and behaviour.

Chaos theory provides a comprehensive and systematic treatment of random-like
complex dynamics in nonlinear systems arising from a broad spectrum of
disciplines. Generally, a chaotic signal is defined in terms of the properties of exhibiting  short-term predictability and long-term unpredictability, something   which is a consequence of the strong dependence on the initial conditions. Therefore, given a temporal signal which describes a quantity of interest in the system, methodologies like
nonlinear forecasting can be used to quantify many important invariants, such
as the fractal dimension, Lyapunov exponents, and entropies for irregular
temporal evolutions yielding a physical description of the corresponding
dynamical structures and predictability properties of the system (see e.g.~\cite{Casdagli1989,Abarbanel1993,Baek_etal_2004}), with many important
applications in e.g.~weather prediction~\cite{Patil_etal_2001,Slingo13122011}
and transistor circuits~\cite{Hanias_etal_2010}.

Likewise, another related question of interest is whether the particular time
series can be distinguished from being stochastic or
chaotic~\cite{Sugihara1990,CAZELLES1992,Tsonis1992}. We note that stochastic effects are ubiquitous
in many systems (see e.g.~\cite{noisy_KS,Nicoli_PRE,Marc_PRL,Marc_EJM} for a
noisy version of the KS equation) and they may be present in any
low-dimensional description of a nonlinear system, as the recent study on
stochastic mode reduction of the KS equation using concepts from evolutionary
renormalization group theory and information entropy
indicates~\cite{Markus2013,Markus2013b}. It is actually in this context that
nonlinear forecasting methodologies become a valuable tool as the more
traditional time-series analysis techniques, such as computing the power
spectral density (PSD) or looking at the trajectories in the phase space,
cannot distinguish between the two; it is noteworthy that both
high-dimensional chaos and stochastic processes produce completely scattered
plots in the phase space. Hence, providing a reliable and systematic
technique that can answer these questions has received considerable attention
over the
years~\cite{Guckenheimer1982,Gao_etal_2006,Rosso_etal_2007,Zunino_etal_2012}.

In this work, we undertake a detailed study of the characteristics of the
dynamical state of the gKS equation in a wide range of the parameter $\delta$
by making use of a time-series analysis based on chaos theory, and in
particular a nonlinear forecasting method to extract the short-term
predictability and long-term unpredictability characteristics of the gKS
solution. We consider in particular two types of temporal signals, a local
quantity, i.e.~a single point in the system, and an appropriate global
measure in the sense that is a spatially averaged quantity. We then apply
several statistical and time-series methods, some of which are standard in
chaos theory, but they have not been used to analyze the gKS solution,
including the Lorenz return map and a nonlinear forecasting method. The
latter in particular has recently been exploited in the context of a wide
spectrum of chaotic systems such as experiments in propagating flame front
instabilities induced by buoyancy/swirl coupling~\cite{Gotoda2010} and
thermoacoustic combustion instabilities in a gas-turbine model
combustor~\cite{Gotoda2011}. Other examples include numerical solutions of
sets of nonlinear differential equations for modeling intermittent
Rayleigh-B\'enard convection subjected to magnetic force \cite{Gotoda2013},
and modeling flame front instabilities induced by radiative heat loss
\cite{Gotoda2012} with emphasis on the treatment of irregular temporal time
series data. This forecasting method is an appropriate extension of the
algorithm proposed by Sugihara and May~\cite{Sugihara1990}, in the sense that
it allows us to quantify the short-term predictability and long-term
unpredictability as a function of the prediction time by taking into account
the update of the library data in the phase space.
This is crucial in order
to understand the predictability properties of the system.

We find that as the parameter $\delta$ is increased both local and global
quantities undergo a transition from high-dimensional chaos, characterized by
scattered graphs in the phase space and a very short predictability time, to
a periodic solution which can be predicted up to long times. At moderate
values of $\delta$ the solution exhibits low-dimensional chaos and is
characterized by a critical predictability time that we directly relate to
the dispersion parameter $\delta$. We also compare the results from the gKS
equation to pure stochastic processes like fractional Brownian motion to
infer conditions under which the methodology can distinguish between the gKS
signal from pure stochastic signals.

Section \ref{Sec : ST gKS} presents the different dynamical states observed
in the gKS equation as the dispersion parameter is increased and we define
there the two temporal signals mentioned earlier. Section~\ref{Sec:NonLinear
F} briefly details the nonlinear forecasting method and Sec.~\ref{Sec : Stoch
vs chaos} outlines the difference between a pure stochastic process and a
deterministic chaotic signal. The results for the local and global analysis
are presented in Sec.~\ref{Sec:Local}, Sec.~\ref{Sec:global}, and
Sec.~\ref{Sec:Ent}. We finally conclude in Sec.~\ref{Sec:conclusions}.

\section{Spatiotemporal dynamics of the \protect{g}KS equation}
\label{Sec : ST gKS}
Consider the gKS equation (\ref{Eq:gKS}) in a domain $[-L,L]$, i.e.~of
size $2L$, with periodic boundary conditions and discretized into $N_L$
points. We choose random initial conditions $u(x,0)=\xi(x)$ where $\xi(x)$ is
a spatial white noise with $\langle\xi(x)\xi(x')\rangle=2\delta(x-x')$. The
numerical solution $u(x,t)$ is obtained by adopting a pseudo-spectral method
for the spatial derivatives that uses the Fast Fourier Transform (FFT) to
transform the solution to Fourier space. The nonlinear terms are evaluated in
real space and transformed back to Fourier space by using the inverse FFT.
The solution is then propagated in time by making use of a modified
fourth-order exponential time-differencing--time-stepping Runge-Kutta
(ETDRK4) scheme~\cite{Kassam2005}, which is very stable for stiff
equations~\cite{Shen2011}. It has been reported in previous
studies~\cite{Elder1997,Wittenberg1999} that for sufficiently large $L$ many
modes become active and the transition to spatiotemporal chaos in the KS
equation becomes independent of $L$. On this basis we choose $L=500$ and
{$N_L$ = 5000. It is also important to emphasize that by taking a large
number of discretised points (in our case $5000$) we ensure the probability
density function of the random initial conditions converges to a Normal
distribution of mean zero and variance one so that starting the computations
with different random initial conditions has little influence on the dynamic
solutions. Note that solutions with different means can be on the other hand
mapped to each other via a Galilean transformation (see
e.g.~\cite{Tseluiko19092012} for the gKS equation and
\cite{Coullet_PRL,Malomed1992} for other systems with this type of symmetry).
We also choose the value $\nu=1$ and we vary the dispersion parameter
$\delta$ from $0$ to $3$. Note also that to improve the numerical
precision, all numerical results, when necessary, have been obtained after
averaging over 10 different realizations of the initial conditions.
\begin{figure}
\centering
\includegraphics[width=0.48\textwidth]{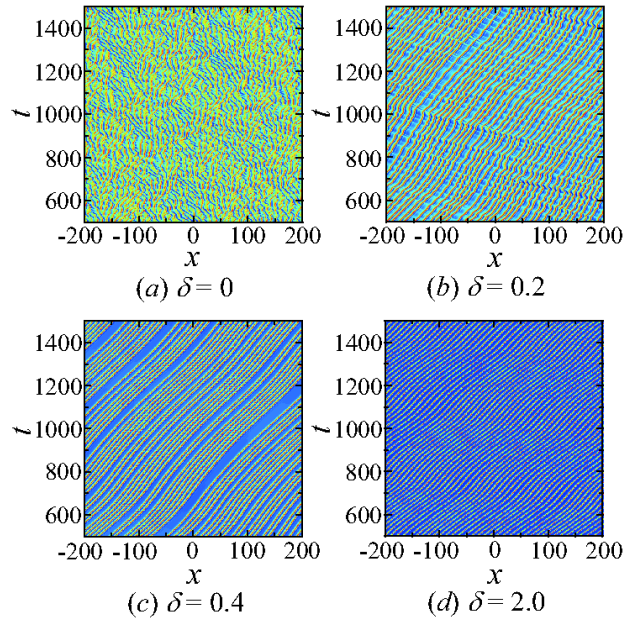}
\caption{Spatiotemporal patterns of the gKS equation for the spatial and temporal
regions of $x\in\left[-200,200\right]$ and $t\in\left[500,1500\right]$,
respectively, and for different values of $\delta$. Different colours represent the magnitude of $u$ with high and low values in yellow and blue,
respectively.}\label{fig:1 patterns}
\end{figure}

Figure \ref{fig:1 patterns} depicts spatiotemporal patterns of the gKS
equation for different values of $\delta$. Clearly, for $\delta = 0$, i.e.,
for the KS solution, the well-known spatiotemporal chaos is formed. As we
increase $\delta$ up to $0.2$ localized coherent structures (solitary pulses)
start to appear but with complicated chaotic dynamics due to their continuous
interaction. Regularly-arrayed pulses with unsteady interpulse distances
become predominant at $\delta$ = 0.4, and for sufficiently large $\delta$,
i.e., in the dispersion-dominated regime, strongly regular structures with
nearly constant interpulse distances are observed. These basic localized
structures were firstly reported by Kawahara~\cite{Kawahara1983}, and their
dynamics and interactions have been extensively studied ever
since~\cite{Balmforth1995,Chang1995,Ei1994,Duprat2009,Tse_PRO,Tse2010a,Tseluiko19092012}.

\subsection{Definition of local and global time series}
\begin{figure}
\centering
\includegraphics[width=0.48\textwidth]{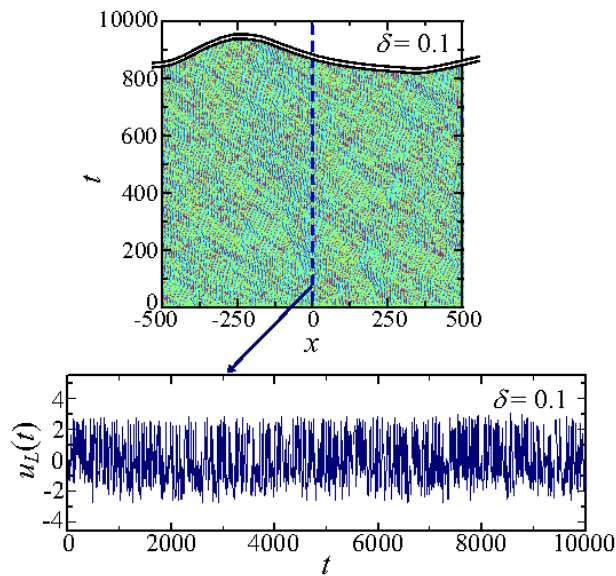}
\caption{Extraction of the temporal evolutions of $u(x,t)$ at $x=0$ which defines the local magnitude $u_L(t)$ for the 1D gKS equation. Different colours represent the magnitude of $u$ with high values (yellow) and low values (blue).} \label{fig:2 Temp Evolution}
\end{figure}
We define here the magnitudes of interest which will be analyzed later on by
means of several time-series tools to characterize the dynamics of the gKS
solution as function of $\delta$. We distinguish between a local quantity,
i.e.~one that represents the dynamics of a single arbitrary point of the
whole domain, and a global quantity as a spatially averaged measure that
gives information of the global dynamics. We hence consider the temporal
evolution of the gKS solution $u(x,t)$ located at $x=0$ and define our local
time signal as:
\begin{equation}\label{Eq:local Def}
u_L(t)\equiv u(0,t).
\end{equation}
Note that this is actually a standard local quantity which has been used  for example  to
characterize the dynamics of Turing patterns in a reaction-diffusion system~\cite{Aragon2012}.
Figure \ref{fig:2 Temp Evolution} shows the temporal evolution of $u_L$ for the case of $\delta=0.1$.

On the other hand, we  define our global measure as the second moment of the gKS solution:
\begin{equation}\label{Eq:global Def}
u_G(t)\equiv \frac{1}{2L}\int_{-L}^{L}[u(x,t)-\overline{u}(t)]^{2}\mathrm{d}x,
\end{equation}
where the overbar denotes spatial average. This quantity has been considered
in other settings, for example in the nonlocal gKS equation describing the
interface evolution of conducting falling films~\cite{Tseluiko2010}. Our
study will be based on how the dispersion (controlled via the parameter
$\delta$) affects the dynamics of both the local $u_L$ and global $u_G$
quantities.

\section{Nonlinear forecasting method}
\label{Sec:NonLinear F}
To study the predictability properties of the gKS equation we adopt a
nonlinear forecasting methodology which is an appropriately extended version
of the algorithm proposed by Sugihara and May~\cite{Sugihara1990}, and it has
been recently applied successfully in a wide spectrum of physical systems,
from thermo-acoustic combustion instability~\cite{Gotoda2011} to flame front
instability induced by radiative heat loss~\cite{Gotoda2012} and intermittent
behavior of Raylegh-B\'enard convection subjected to magnetic
force~\cite{Gotoda2013}. The method proceeds as follows. Given a temporal
signal, say $u(t)$, for $t\in [0,T_f]$, we divide it into two intervals,
namely $t\in[0,t_L]$ and $t\in(t_L,T_f]$, corresponding to a library and
reference set of data, respectively. The values from the library data are
used to predict the temporal evolution of $u(t)$ for $t>t_L$ which are then
compared to the reference data.

We construct the phase space from the time series data based on Takens'
embedding theorem~\cite{Takens1981} by considering time-delayed coordinates,
i.e.~a vector in the phase space is given as:
\begin{equation}
{\bf U}(t_i)=(u(t_i), u(t_i-\tau), ... , u(t_i-(D-1)\tau)),
\end{equation}
where $i=1,\dots,n$ with $n$ the number of values in the data set,
$D$ is the dimension of the phase space, and $\tau$ is a
lag time which is estimated by making use of the mutual information proposed
by~\cite{Fraser1986}, in a similar way as it was also done
in Refs.~\cite{Gotoda2012,Gotoda2013}. In this study we set $D=3$ as we find it to be the optimal value which compromises numerical convergence (i.e.~results do not change for larger values of $D$) and computational time (large values of $D$ become computationally expensive). We define ${\bf
U}_f\equiv{\bf U}(t_f)$ as the final point of a trajectory in the phase
space, and ${\bf U}_k$, for $k=1\dots K$, its $K$ neighboring vectors which are
searched from all data points in the phase space.  The future value
corresponding to ${\bf U}_k$ after some time $T=N\Delta t$ is denoted as
$u(t_k+T)$, where $\Delta t$ is the sampling time of $u(t)$ and $N$ an
arbitrary integer. The predicted value $\check{u}(t_f+T)$ from ${\bf U}_f$ is
then obtained by a nonlinearly weighted sum of the library data $u(t_k+T)$
given  as:
\begin{equation*}
\check{u}(t_f+T)=\frac{\sum_{k=1}^{K}u(t_{k}+T)\mathrm{e}^{-d_{k}}}{\sum_{k=1}^{K}\mathrm{e}^{-d_{k}}},
\end{equation*}
where $d_{k}=\left\Vert{\bf U}(t_{f})-{\bf U} (t_{k})\right\Vert$. A comparison between the predicted value $\check{u}(t_f+T)$ and the reference value $u(t_f+T)$ is then quantified by looking at the standard correlation factor which is defined as:
\begin{equation}\label{Eq:C coeff}
C=\frac{\mathbb{E}[u(t)\check{u}(t)]}{\sigma_u\sigma_{\check{u}}},
\end{equation}
where $\mathbb{E}[u(t)\check{u}(t)]$ represents the covariance between both
signals, and $\sigma_u$ and $\sigma_{\check{u}}$ are the standard deviation
of $u$ and $\check{u}$, respectively. It should be noted that our nonlinear
forecasting method was developed in Refs.
\cite{Gotoda2011,Gotoda2012,Gotoda2013}, where the temporal evolutions of $u$
are predicted by updating the library data and so we are able to capture the
determinism of the system (with keeping the size of the updated library data
constant). This point is actually the difference with the Sugihara-May
algorithm~\cite{Sugihara1990} where the library data is not updated. In this
sense, the correlation coefficient $C$ is estimated in terms of the duration
$t_{P}$ of the actual temporal evolutions of $u$ added to the library data
which is something that has not been proposed before for extracting the
short-term predictability and long-term unpredictability characteristics of
chaotic dynamics.
\section{Chaotic versus stochastic behavior}
\label{Sec : Stoch vs chaos}
Chaotic signals are characterized by exhibiting a high sensitivity on the
initial conditions which in turn induces a decay of predictability in time.
In particular, they exhibit short-term predictability and long-term
unpredictability of the trajectories in the phase space, something which is
referred to as orbital instability. A natural question of fundamental
interest is then, given a time signal how can one distinguish between a pure
deterministic chaotic signal from a stochastic process. Although in this work
we focus on the deterministic gKS equation, we also analyze a stochastic signal by
means of the nonlinear forecasting method which will be
used in the subsequent sections for comparison with pure chaotic signals.

Consider the fractional Brownian motion which is a Gaussian process
$B_\alpha(t)$ with zero mean, i.e.~$\mathbb{E}[B_\alpha(t)]=0$,  and a time
dependent variance and covariance which are given as:
\begin{eqnarray}
\mathbb{E}[B_\alpha(t)B_\alpha(t)] & = & t^{\alpha-1},  \\
\mathbb{E}[B_\alpha(t)B_\alpha(s)] & = & (t^{\alpha-1}+s^{\alpha-1}-\vert s-t\vert^{\alpha-1})/2,
\end{eqnarray}
respectively, where $\alpha$ is the scaling exponent~\cite{Mandelbrot_1968}.
This is a well-known stochastic process model, the PSD of which exhibits a
power-law behaviour of the type $1/f^{\alpha}$, describing irregular behavior in many areas of science and
engineering~\cite{Biagini2008}. On the other hand, the increment process
defined as $\Delta B_\alpha\equiv B_\alpha(t_{i+1})-B_\alpha(t_{i})$ is
fractional Gaussian noise. The particular case of $\alpha = 5/3$ is referred
to as a Gaussian Kolmogorov signal, in reference to the same type of PSD
observed in fully-developed turbulence~\cite[see
e.g.][]{Vergassola1993,Katul2001}.
For $\alpha = 2$, it corresponds to Brownian
motion, and the increment process is white noise.

To demonstrate the ability of the nonlinear forecasting methodology proposed
in the preceding section for distinguishing chaos from stochastic process, we
investigate the variations of the correlation coefficient $C$ in terms of the
predicted time $t_{P}$ for the two representative cases of Brownian ($\alpha
= 2$) and fractional Brownian motion ($\alpha = 5/3$) and the corresponding
increment processes. We hence take $u(t)= B_\alpha(t)$ in Eq.~(\ref{Eq:C coeff})
and compute the corresponding correlation coefficient for the two aforementioned
values of $\alpha$. The results for both cases are shown in Fig.~\ref{Fig:
Stochastic analysis}. We observe that $C(t_P=1)\sim 1$ in both cases and it
decreases with $t_{P}$. This means that the one-step-ahead
short-term prediction with high accuracy is feasible for Brownian/fractional
Brownian motion. However, when we compute the increment process $\Delta
B_\alpha$ the correlation factor is almost zero regardless of $t_{P}$,
indicating that it cannot be predicted. As we will show in the following
section, this is a distinctive feature of a pure stochastic process and will
allow us to distinguish it from the deterministic chaos of the gKS equation.

\begin{figure}
\includegraphics[width=0.95\textwidth] {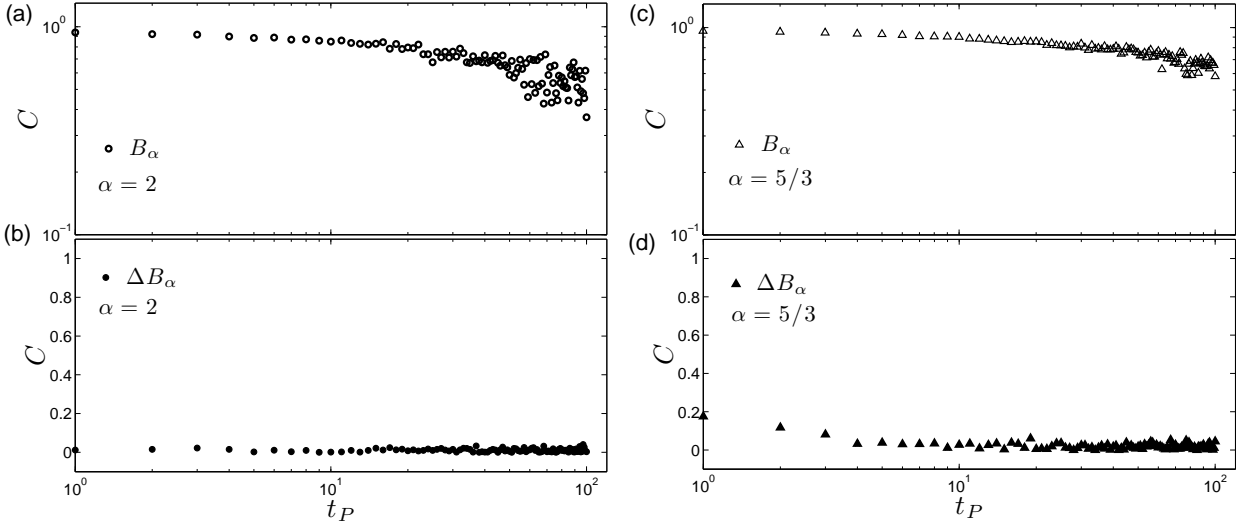}
\caption{Correlation coefficient $C(t_P)$ for (a) Brownian motion $B_\alpha(t)$ with $\alpha=2$ and the
corresponding increment $\Delta B_\alpha(t)$ (b); and (c)
fractional Brownian motion $B_\alpha(t)$ with $\alpha=5/3$ and the
corresponding increment $\Delta B_\alpha(t)$ (d). Panels (a) and (c) are plots
in log-log scale and panels (b) and (d) in semi-log scale.} \label{Fig:
Stochastic analysis}
\end{figure}

\section{Local analysis}
\label{Sec:Local}
\subsection{Transition from chaotic to periodic solutions in the gKS equation}
We start by studying how the dynamical state of the gKS solution undergoes
the transition from a chaotic behavior, observed at low values of $\delta$,
to a pulsating periodic dynamics which appears at sufficiently large values
of $\delta$. To this end, we are going to perform a time-series analysis.
\begin{figure*}
\centering
\includegraphics[width=0.8\textwidth] {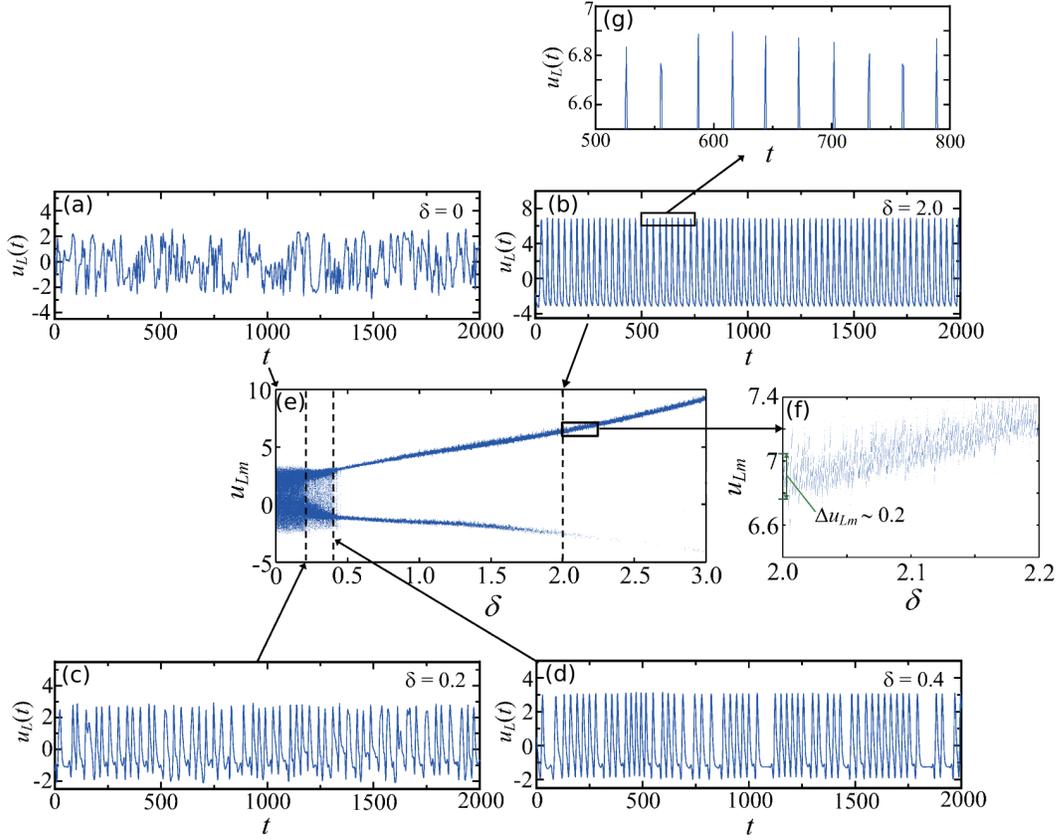}
\caption{(a-d) Temporal evolutions of $u_L(t)$ for different values of
$\delta$. (e) Bifurcation diagram of $u_{Lm}$. Panels (f) and (g) show a
magnification of the selected area in panels (e) and (b), respectively.}
\label{fig:4 bifurcation}
\end{figure*}

\subsubsection{Bifurcation diagram.}We look at the bifurcation diagram of the gKS equation by plotting the local maxima of the temporal signal
$u_L(t)$ which we define as the values $u_{Lm}$ that satisfy the following
condition:
\begin{equation}
u_{Lm} = \{u_L(t)\ :\ \dot{u}_L(t)=0,\ \ddot{u}_L(t)<0\}.
\end{equation}
Temporal evolutions of $u_L(t)$ and the corresponding bifurcation diagram for
different values of $\delta$ are shown in Fig.~\ref{fig:4 bifurcation}. We
note that a repeated oscillation of $u_L(t)$ is displayed as a single point
in the bifurcation diagram. We observe that $u_L(t)$ exhibits highly
irregular fluctuations at $\delta=0$, which give rise in turn to the
formation of uniformly distributed values of $u_{Lm}$ in the bifurcation
diagram. For $\delta=0.2$, $u_L(t)$ starts to exhibit pulsating time
fluctuations, and a hollow region appears in the bifurcation diagram around
the values of $0\le u_{Lm}\le 2$. For $\delta \gtrsim 0.4$, these pulsating
oscillations become more clear with temporal structures separated by
irregular times, which leads to the appearance of two dominant branches in
the bifurcation diagram. At this point the amplitude of the oscillating
signal $u_L(t)$ is notably amplified as $\delta$ is increased which induces a
gradual increase of the width separating the two branches in the bifurcation
diagram.

In the dispersion-dominated region of $\delta\ge 2$, only the upper
dominant branch persists, indicating the formation of strongly periodic
pulsed oscillations. Note however that, as shown in the top right panel of Fig.~\ref{fig:4 bifurcation},
the pulsed amplitude is oscillating in time which is a consequence of the
weak interaction between pulses. For larger values of $\delta$
such interactions would become weaker and a pure periodic signal would be recovered.
It is important to point out that the main relevant change in the bifurcation
diagram occurs at $\delta \sim 0.4$ where there is an indication of the onset
of interior crisis. Several kind of crisis
phenomena in the  KS equation were found by~\cite{Chian2002,Rempel2003},
but there are no previous studies reporting the existence of such interior crisis in
the temporal evolution of $u_L(t)$ for the gKS equation.

\subsubsection{Power spectral density.} We analyze next the PSD of the signal $u_L(t)$ for different values of $\delta$ which is defined as
\begin{equation}\label{Eq:PSD def}
S_u(f)=\langle\vert\hat{u}_L(f)\vert^2\rangle,
\end{equation}
where $\hat{u}_L(f)$ represents the Fourier transform in the frequency domain
of $u_L(t)$ which is defined as $\hat{u}_L(f)=
(1/\sqrt{T})\sum_{t}[u_L(t)-\overline{u}_L]\exp{(ift)}$, where $T$ is the
time interval of observation and $\overline{u}_L$ is the mean value of the
signal $u_L(t)$ over $T$. The symbols $\langle\dots\rangle$ in
Eq.~(\ref{Eq:PSD def}) denote average over different time intervals - we note
that we are assuming statistically stationary solutions. Figure \ref{fig:5
PSD} shows that for $\delta\lesssim 0.2$ the PSD consists of a
continuous distribution region at low frequencies followed by an exponential
decay at higher frequencies. It is important to note that an exponential
behaviour for the PSD is usually associated with deterministic chaotic
systems (see e.g.~\cite{Maggs_2011} and references therein) which indicates
that the gKS solution is clearly chaotic for small values of $\delta$. For
$\delta=0.4$ multiple peaks appear ($2f_{m}$, $3f_{m}$, $4f_{m}\dots$) with
basic frequency $f_{m}=0.035$ corresponding to the temporal pulsating
structures of $u_L(t)$ observed in Fig.~\ref{fig:4 bifurcation}, and indicate
the multiple-periodicity nature of the signal. The appearance of these peaks
becomes more prominent in the dispersion-dominated region of $\delta = 2$.
\begin{figure}
\centering
\includegraphics[width=0.95\textwidth]{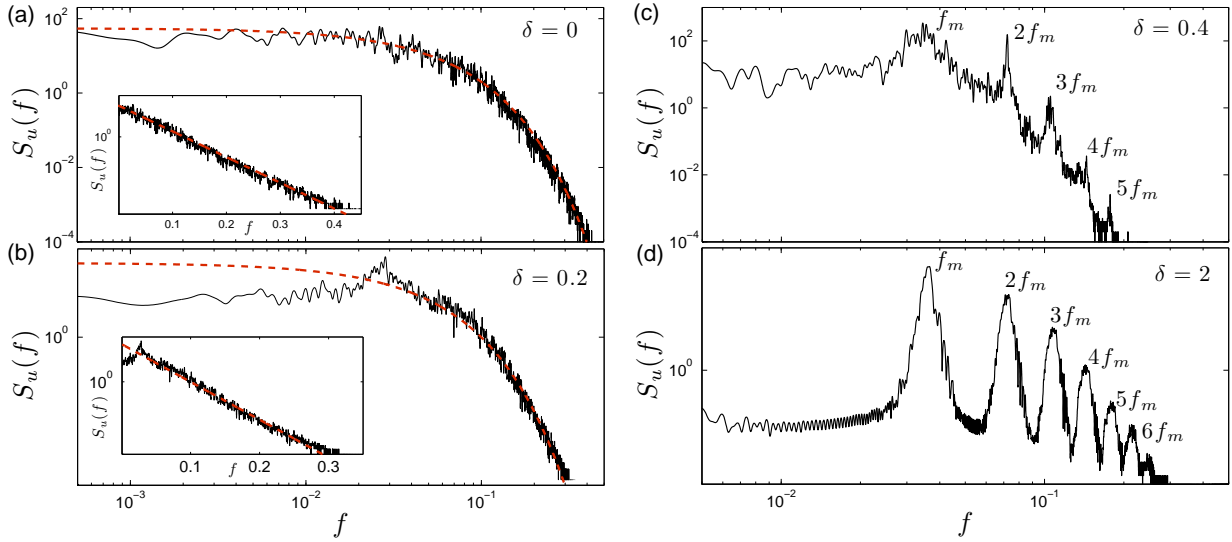}
\caption{Log-log plot of the PSD $S_u(f)$ of $u_L(t)$ for different values of $\delta$. The insets in (a) and (b) correspond
to a semi-log plot to show the exponential dependence. The dashed lines correspond
to an exponential function $\exp{(-f\tau_0)}$ with $\tau_0=33$ for $\delta =0$ (a) and $\tau_0=48$ for $\delta =0.2$ (b).}
\label{fig:5 PSD}
\end{figure}

\subsubsection{Dynamical structure of the phase space.} We compute the Lorenz return map which is obtained by
plotting pairs of successive local maximum points, i.e.~we define the $k$th
local maximum as ${u_{Lm,k}}$ and we plot ${u_{Lm,k+1}}$ against
${u_{Lm,k}}$. Figure \ref{fig:6 map}(a) depicts the results for $\delta = 0$,
$0.2$, $0.4$ and $2$. An entirely scattered structure appears for $\delta =
0$, indicating the presence of chaotic behavior. For $\delta = 0.2$ we get a
map which is mainly gathered in three or four regions with scattered
structures, which with increasing $\delta$ up to $0.4$, they progressively
group in four main regions with locally scattered plots. Interestingly, two
of them follow the straight line ${u_{Lm,k+1}}={u_{Lm,k}}$. In the
dispersion-dominated regime with $\delta = 2$, these four points aggregate at
one region which belongs to the line ${u_{Lm,k+1}}={u_{Lm,k}}$. This is
evidence of the formation of strongly periodic pulsating oscillations, but
note that the group located at the top right of the map is still exhibiting a
scattered structure. Note also that the amplitude of $u_{Lm}$ in
Fig.~\ref{fig:4 bifurcation} yields the entirely scattering plots in the
localized region. This indicates that the dynamical state of $u$ at  $\delta
= 2$ is dominated by a strongly periodic pulsed oscillation, but it involves
small fluctuations in the oscillation amplitude (see top right panel of
Fig.~\ref{fig:4 bifurcation}).
\begin{figure}
\centering
\includegraphics[width=0.9\textwidth] {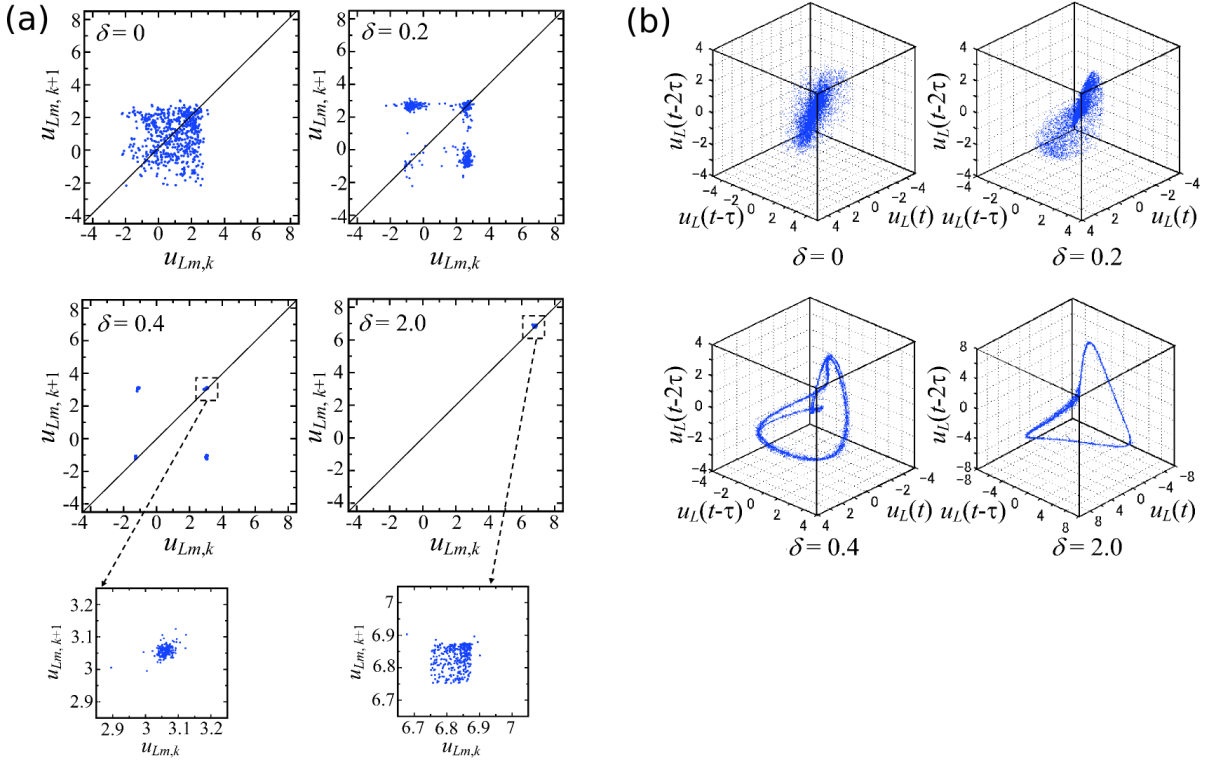}
\caption{(a) Lorenz return map of $u_L(t)$ for different values of $\delta$.
(b) 3D phase space $\left[u_L(t),u_L(t-\tau),u_L(t-2\tau)\right]$ for  $\delta=0$ ($\tau=2$), $\delta=0.2$ ($\tau=2$), $\delta=0.4$ ($\tau=6$), and
$\delta=2$ ($\tau=7$).}
\label{fig:6 map}
\end{figure}

We also look at the three dimensional (3D) phase space which we construct by
considering $\left[u_L(t),u_L(t-\tau),u_L(t-2\tau)\right]$. The results for
different values of $\delta$ are plotted in Fig.~\ref{fig:6 map}(b).
As before, we observe a entirely scattered graph for $\delta = 0$ which fills
the core of the attractor, indicating hence the presence of high-dimensional
chaos. It still remains a scattered structure for $\delta = 0.2$, but is
different from that at $\delta = 0$. The difference between the two cases is
hard to be extracted by the conventional linear analysis such as the PSD (see
Fig. 4). In this sense, drawings of the Lorenz return map and the 3D phase
space are useful for capturing the subtle differences between the dynamical
state of the KS and the gKS equation with small $\delta$. With $\delta$
increasing up to $0.4$ the overall shape of the attractor exhibits the
geometrical structure of a strange attractor which is a consequence of the
occurrence of multiple-periodic pulsed oscillations with irregular time
intervals, indicating hence the presence of low-dimensional chaos.
For $\delta=2$, we observe the emergence of a limit cycle
with some small fluctuations corresponding to the fluctuating amplitude (see
Fig.~\ref{fig:4 bifurcation}).


\subsection{Nonlinear forecasting analysis}
\label{Sec: forecasting analysis}
We apply the methodology described in Sec.~\ref{Sec:NonLinear F} to study the
predictability properties of $u_L(t)$. We divide the time series
$u_L(t)$ into library and reference data by considering the values of $u_L$
up to $t_L = 2500$ to be used as library data and those thereafter as
reference data.

Figure \ref{fig:8 Correlation}(a) depicts the correlation coefficient $C$ as
function of the prediction time $t_{P}$ for different values of $\delta$. We observe
that for the KS equation ($\delta=0$), $C$ at $t_{P}=1$ is nearly unity,
which means that one-step-ahead prediction of the dynamical state is feasible
with high accuracy. However, $C$
decreases rapidly with $t_{P}$ reaching nearly zero at $t_{P}=100$. 
As $\delta$ is increased, we observe that a short plateau of constant $C$
starts to appear which, after a critical predicted time, say $t_{P,c}$, is followed by a rapid
decay the rate of which depends on the particular value of $\delta$, and which can be
approximated by a power-law tail, $C(t_P)\sim t_P^{-\gamma}$, with the exponent $\gamma$ depending
on $\delta$. In particular we find that
\begin{equation}
C(t_P)\sim \left\{
  \begin{array}{l l}
    1 & \quad \text{if $t_P<t_{P,c}$},\\
    t_P^{-\gamma} & \quad \text{if $t_P>t_{P,c}$},
  \end{array} \right.
\end{equation}
with $\gamma = 0.67\pm 0.06$ for $\delta = 0$, $\gamma = 0.55\pm 0.07$ for
$\delta = 0.1$, and $\gamma = 0.41\pm 0.04$ for $\delta \ge 0.2$. Based on
the above relationship, the predicted critical time $t_{P,c}$ is hence
numerically obtained by fitting a power law at long times and a constant
function at short times, and find the point where these two behaviours meet
[see the inlet panel of Fig.~\ref{fig:8 Correlation}(b) for an example].
The fact that the same exponent $\gamma$ is observed for $\delta \ge 0.2$ suggests data of $C$ for different values
of $\delta$ should collapse when plotted against the ratio $t_P/t_{P,c}$. The results are presented in
Fig.~\ref{fig:8 Correlation}(b), where we can observe an excellent data collapse indicating
indeed that for $\delta \ge 0.2$ the exponent $\gamma$ becomes independent of $\delta$.
Note that the appearance of such constant region followed by a rapid decrease is the manifestation
of short-term predictability and long-term unpredictability characteristics
of chaos which are separated by the critical value $t_{P,c}$. It is important to emphasize that even in the case of $\delta=0.4$, where the spatial solution is characterized by localized structures, the
temporal behavior exhibits a chaotic motion which is a consequence of the
irregular interaction between these structures. This regime corresponds to
low-dimensional chaos, or what sometimes is referred to as weak/dissipative
turbulence (in the ``Manneville sense"~\cite{Manneville1985}, as noted in the
Introduction).

In the dispersion-dominated region with $\delta = 2$, the region of constant
$C$ becomes significantly larger followed by a slower decay. We
note that if the dynamical state is periodic oscillations, $C$ is unity
regardless of $t_{P}$ meaning a long-term predictability (see inlet panel of
Fig.~\ref{fig:8 Correlation}(a) for an example of temporal prediction of a sine
wave $\sin{\left(2\pi ft\right)}$ with $f=0.035$).  The gradual decay of the
gradient of $C$ at $\delta = 2$ is attributed to small fluctuations in pulsed
oscillation amplitude extracted by the Lorenz return map, i.e.
uniformly-scattered plots in a single region.
\begin{figure}
\centering
\includegraphics[width=0.92\textwidth] {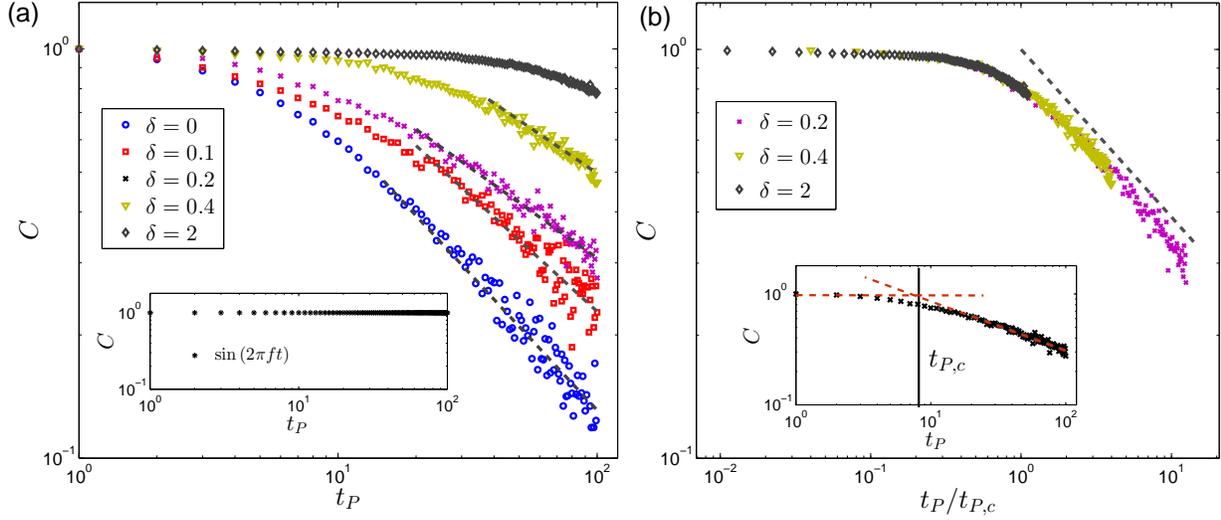}
\caption{(a) Log-log plot of the time correlation coefficient $C(t_P)$ for $u_L(t)$ against the predicted
time $t_{p}$ for different values of $\delta$. The inlet panel corresponds to
the correlation coefficient for a pure periodic function $\sin{(2\pi ft)}$. (b)
Data collapse of $C$ against $t_P/t_{P,c}$ where $t_{P,c}$ is the critical
predicted time after which $C$ starts to decay. The dashed line corresponds to a
decaying power-law tail with exponent $-0.4\pm 0.03$. The inlet panel shows how
the critical predicted time $t_{P,c}$ is defined.}
\label{fig:8 Correlation}
\end{figure}

We also scrutinize the increment process of $u_L$ which we define as $\Delta
u_L\equiv u_L(t_{i+1})-u_L(t_i)$. Figure \ref{fig:9 Increment process} shows
the correlation coefficient $C$ for $\delta = 0$, $0.2$ and $0.4$, where we
observe a similar trend as we did with the results for $u_L$
[cf.~Fig.~\ref{fig:8 Correlation}(a)]. We note that the behavior of $\Delta
u_L$ is completely different to the one observed for the increment process of
the fractional Brownian dynamics (see Fig.~\ref{Fig: Stochastic analysis}).
This indicates that a way of distinguishing between chaos and pure
stochasticity is by looking at the predictability properties of the increment
process, as opposed of looking at the original signal - this is particularly
true for high-dimensional chaos ($\delta=0$) as both signals exhibit a
similar trend for $C(t_P)$.

Based on these results, the short-term predictability and long-term
unpredictability of the gKS equation are summarized in Fig.~\ref{fig:10
Sketch}. As we increase $\delta$, the dynamical state is stabilized and
the gradient of the correlation coefficient against the prediction time,
i.e.~the exponent $\gamma$, decreases. The degree of short-term
predictability can then be roughly estimated by the critical prediction time
$t_{P,c}$. Figure \ref{fig:11 critical bh} shows this time and the exponent
$\gamma$ as function of $\delta$, observing that $t_{P,c}$ gradually
increases for $\delta \ge 0.2$ and $\gamma$ rapidly decreases for $\delta <
0.2$ and remains constant around the value $0.4$ until $\delta \sim 1$, after
which starts slowly decreasing. It should be pointed out however that for
these large values of $\delta$, the power-law behaviour at long times (and in
turn the exponent $\gamma$) is largely affected by cross-over effects: to
find the truly power-law decay we would need to go to longer values of
$t_P$. These three regimes are identified with the existence of
high-dimensional spatiotemporal chaos at low values of $\delta$,
low-dimensional chaos at moderate values, and periodic solutions at large
values. These results demonstrate that the proposed nonlinear forecasting
methodology allows us to fully characterize the dynamics of the system in
terms of its predictability properties.

\begin{figure}
\centering
\includegraphics[width=0.45\textwidth] {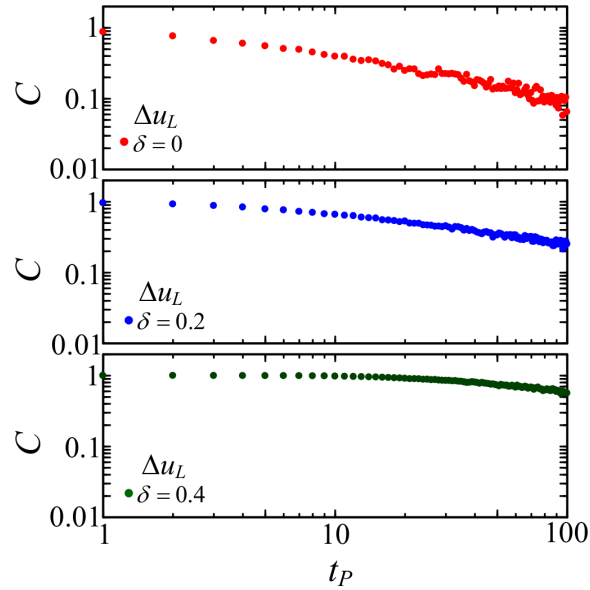}
\caption{Correlation coefficient $C(t_P)$ for the increment process
$\Delta u_L(t)$ and for different $\delta$.}
\label{fig:9 Increment process}
\end{figure}

\begin{figure}
\centering
\includegraphics[width=0.45\textwidth] {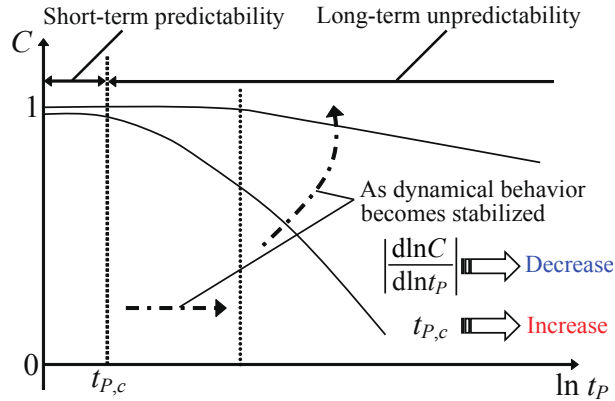}
\caption{Schematic representation of short-term predictability and
long-term unpredictability characteristics.}
\label{fig:10 Sketch}
\end{figure}

\begin{figure}
\centering
\includegraphics[width=0.45\textwidth] {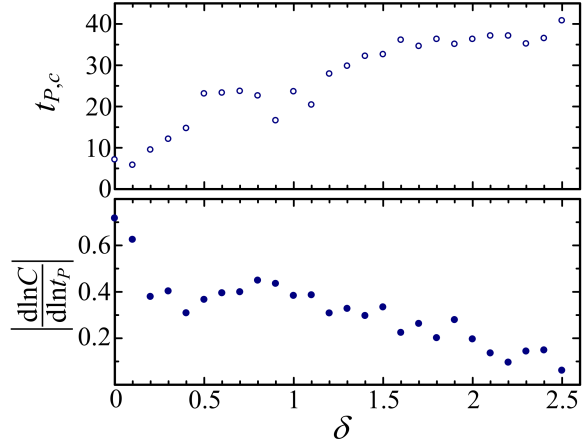}
\caption{Variations in $t_{P,c}$ and
$\gamma=\vert\mathrm{d}\ln C/\mathrm{d}\ln t_{P}\vert$ for
$u_L(t)$ as a function of $\delta$. Typical errors are $3$ and $0.03$ for
$t_{P,c}$ and $\gamma$, respectively. }
\label{fig:11 critical bh}
\end{figure}

\section{Global analysis}
\label{Sec:global}

We repeat the same analysis as in Sec.~\ref{Sec:Local} but considering the
global magnitude $u_G(t)$ as defined via the second moment in
Eq.~(\ref{Eq:global Def}). Figure \ref{fig:12:global u}(a) depicts the temporal
evolutions of $u_G(t)$ for different $\delta$. Irregular fluctuations are
clearly formed for $\delta\lesssim 0.2$ and we now begin to observe some
pulsating oscillations with irregular time interval separations at
$\delta=0.4$ which become more regular at $\delta=2$. It is important to
emphasize that in contrast to what we observed for the local magnitude $u_L$
[cf.~Fig.~\ref{fig:4 bifurcation}], these pulsating oscillations are
superimposed with a long wave oscillation of irregular frequency.
\begin{figure}
\centering
\includegraphics[width=0.9\textwidth] {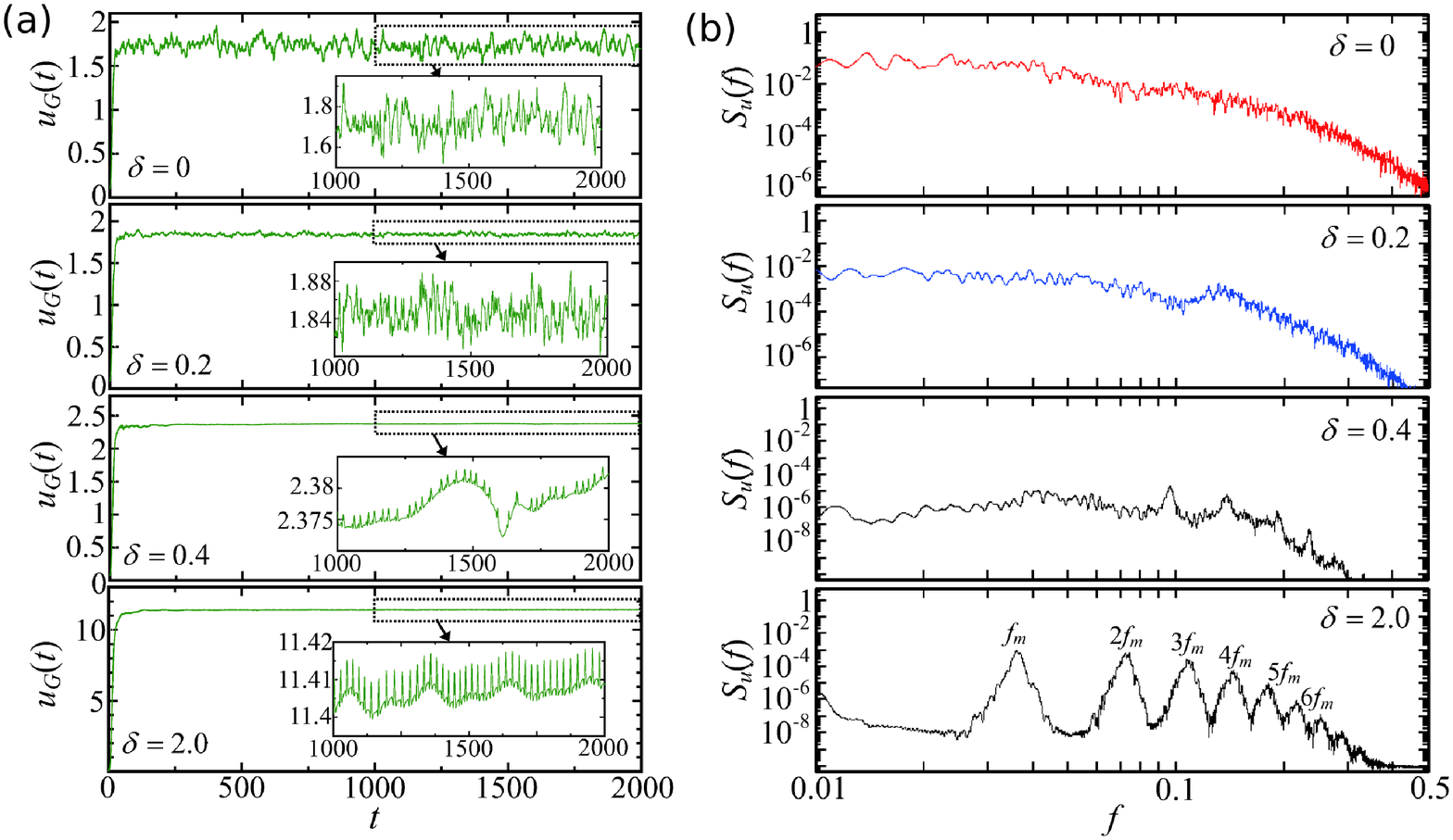}
\caption{(a) Temporal evolutions of $u_G(t)$ for different $\delta$. (b) PSD of the global magnitude $u_G(t)$ for different $\delta$.}
\label{fig:12:global u}
\end{figure}

We compute the PSD of $u_G(t)$ observing a similar behavior as for the local
magnitude $u_L$ when $\delta$ is increased [see Fig.~\ref{fig:12:global
u}(b)]. We also look at the Lorenz return map and the 3D phase space
observing that now the transition from a completely scattered graph to
structured graphs appears to be at larger $\delta$ values (see
Fig.~\ref{fig:14 Lorenz global}). Figure \ref{fig:16 C global}   shows the
correlation coefficient $C$ for the signal $u_G$ and for the increment
process $\Delta u_G\equiv u_G(t_{i+1})-u_G(t_i)$, respectively. As
before, we observe an initial regime of short-term predictability followed by
a power-law decay of $C$ indicating the long-term unpredictability which is
decreased as we increase $\delta$.

We note that the formation of periodic pulsed oscillations which are now
superimposed on a long wave of irregular frequency (see
Fig.~\ref{fig:12:global u} for $\delta=0.4$) makes the signal $u_G$ more
chaotic as compared to $u_L$, in the sense that now the transition to
long-term unpredictability appears before than in the case of $u_L$. This is
more clear if we plot the critical predicted time $t_{P,c}$ against $\delta$, observing that now it increases until $\delta\sim 1.1$
after which it remains constant until $\delta=2.0$ (cf.~Fig.~\ref{fig:18
critical global}). As for the gradient $\vert\mathrm{d}\ln C/ \mathrm{d}\ln{t}_{P}\vert$, we observe that it initially decreases for $\delta\ge 0.2$ after which it remains constant until $\delta \sim 2$ for which it starts decreasing again. It is noteworthy that the critical predicted time is
shorter than that for the local signal indicating hence that the global
quantity appears to be less predictable.

\section{Permutation entropy} \label{Sec:Ent}
Finally, we present an alternative analysis of both the local and global
signals, $u_L$ and $u_G$, respectively, based on the concept of permutation
entropy, which is an invariant measure of the complexity of the
dynamics~\cite{Bandt2002}, and has been shown to be a good candidate to
e.g.~quantifying the dynamics in combustion instability~\cite{Gotoda2012_2}.
This analysis complements the previous results and provides more information
about the difference between the magnitudes of the local and global signals.
A modified version of the permutation entropy has also been recently proposed
to study heartbeat dynamics~\cite{Bian2012}.

The permutation entropy $h_p$ can be estimated by quantifying the degree of
randomness from a sequence of ranks in the values of a time series.
In particular, given a sequence with embedding dimension $D_e$, we index all possible permutations ($D_e!$) of order $D_e\ge 2$ as $\pi$. Each permutation represents a coarse-grained pattern of the dynamics when the sequence consists of $D_e$ successive data points taken from a time series.  We first count the number of permutations denoted as $q(\pi)$ for all vectors ${\bf X}(t)=(u(t),u(t+1),\dots,u(t+D_e-1))$ consisting of sequences of order $D_e$. We then calculate the relative frequency for each permutation to obtain $p_e(\pi)=q(\pi)/(N-(D_e+1))$. We set $D_e=7$  in this study, and following the definition of Shannon entropy, $h_p$ is then calculated as:
\begin{equation}
h_{\mathit{p}} =\frac{-\sum\limits_{\pi}{p_{\mathit{e}}({\pi})}{\log_2}{p_{\mathit{e}}({\pi})}}{\log_2{D_e!}},
\end{equation}
where $\log_{2}D_e!$ corresponds to the maximum permutation entropy and is
used for normalization so that $0\le h_p\le 1$. Note that with this definition
the lower bound ($h_p=0$) corresponds to a monotonically increasing or decreasing
process, while the upper bound ($h_p = 1$) corresponds to a completely random process.

Figure 19 shows the results of $h_p$ for both $u_L$ and $u_G$ as we change
the values of $\delta$ from the high-dimensional chaos region ($\delta = 0$) to the
low-dimensional chaos region ($\delta = 0.4$). We observe that $h_p$ decreases for
both $u_L$ and $u_G$ with increasing $\delta$, as a consequence of the dispersion
mechanism that arrests chaotic behavior. It is important to note that $h_p$ is
significantly larger for the case of $u_G$, indicating that the dynamics of the
global measure is more difficult to analyse than the corresponding local one, something which is a consequence of the averaging procedure.

\begin{figure}
\centering
\includegraphics[width=0.95\textwidth] {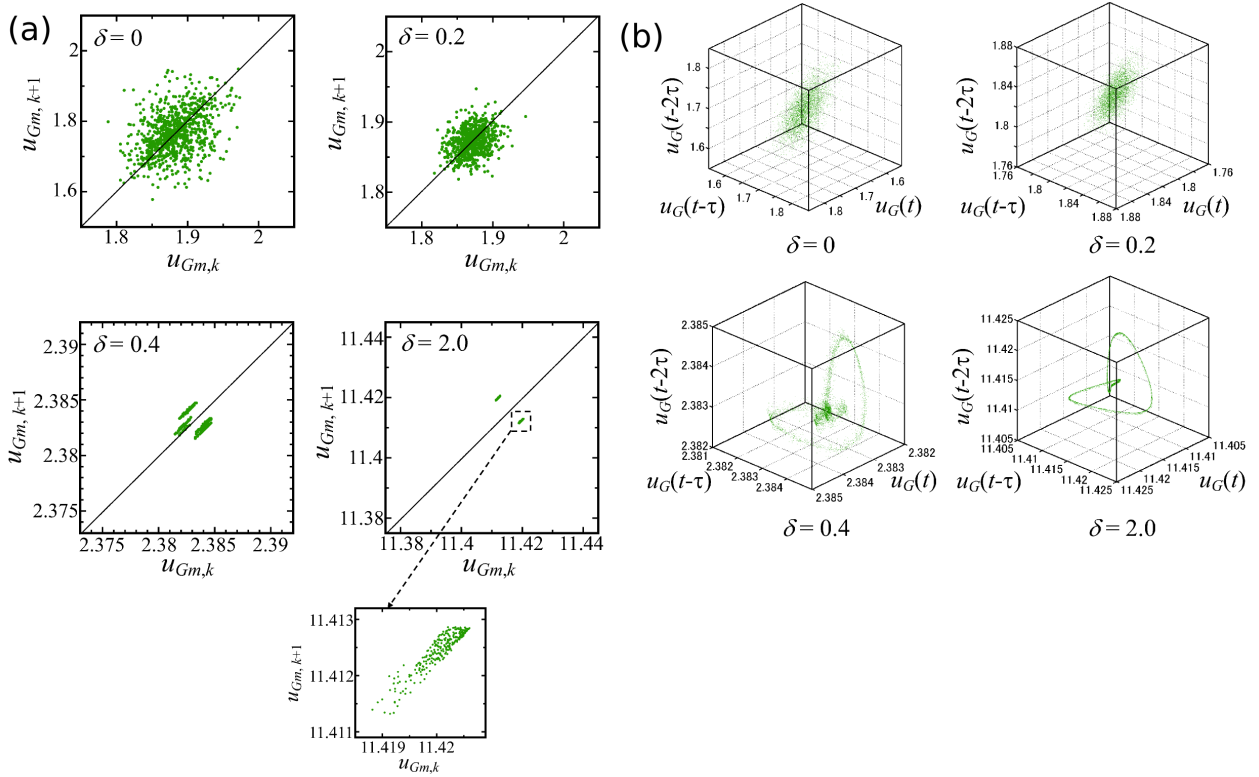}
\caption{(a) Lorenz return map of $u_G(t)$ for different $\delta$. (b) 3D phase space of the global magnitude $u_G(t)$ for  $\delta=0$ ($\tau=2$), $\delta=0.2$ ($\tau=2$), $\delta=0.4$ ($\tau=3$), and
$\delta=2$ ($\tau=8$).}
\label{fig:14 Lorenz global}
\end{figure}

\begin{figure}
\centering
\includegraphics[width=0.95\textwidth] {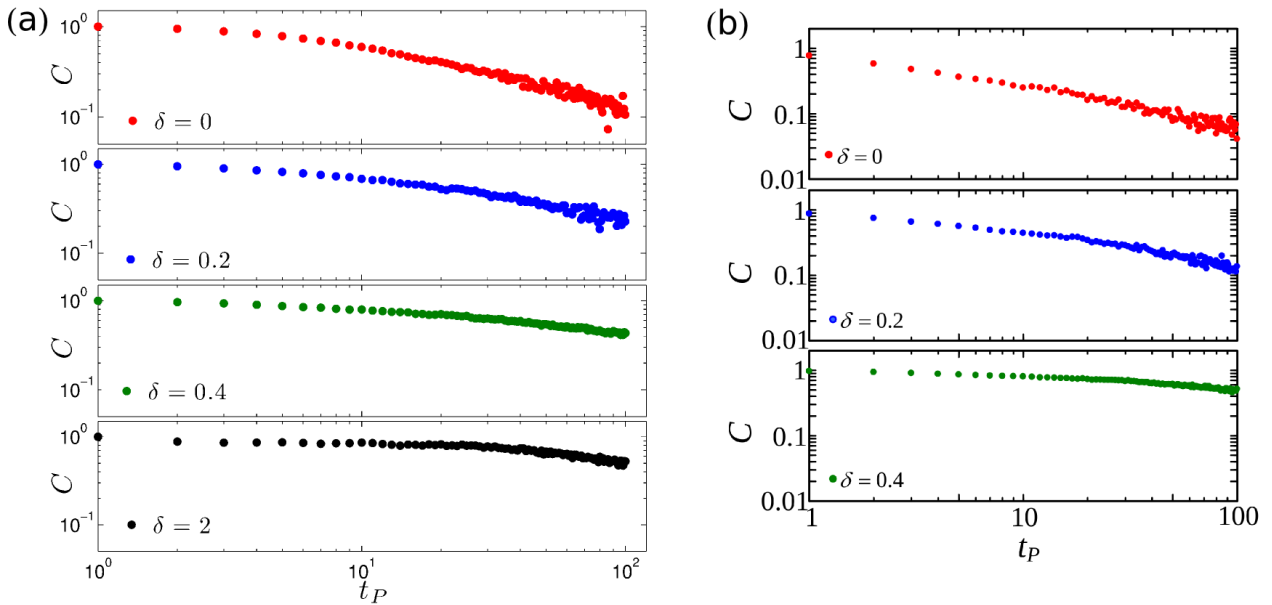}
\caption{(a) Correlation coefficient $C(t_P)$ of $u_G(t)$ (a) and
$\Delta u_G(t)$ (b) for different values of $\delta$.}
\label{fig:16 C global}
\end{figure}
\begin{figure}
\centering
\includegraphics[width=0.45\textwidth] {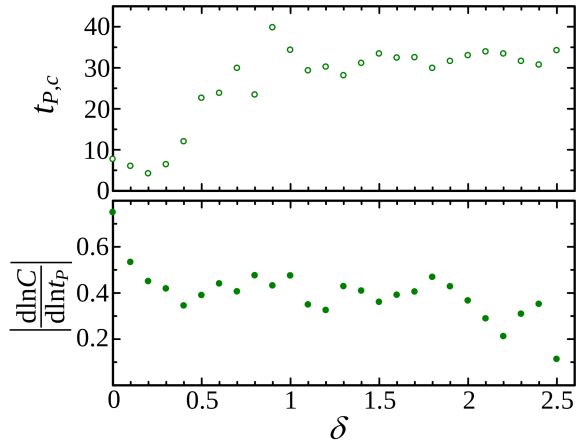}
\caption{Variations of $t_{P,c}$ and  $\gamma=\vert \mathrm{d}\ln C/\mathrm{d}\ln t_{P}\vert $ for $u_G(t)$
as a function of $\delta$. Typical errors are $4$ and $0.03$ for
$t_{P,c}$ and $\gamma$, respectively.}
\label{fig:18 critical global}
\end{figure}

\begin{figure}
\centering
\includegraphics[width=0.43\textwidth] {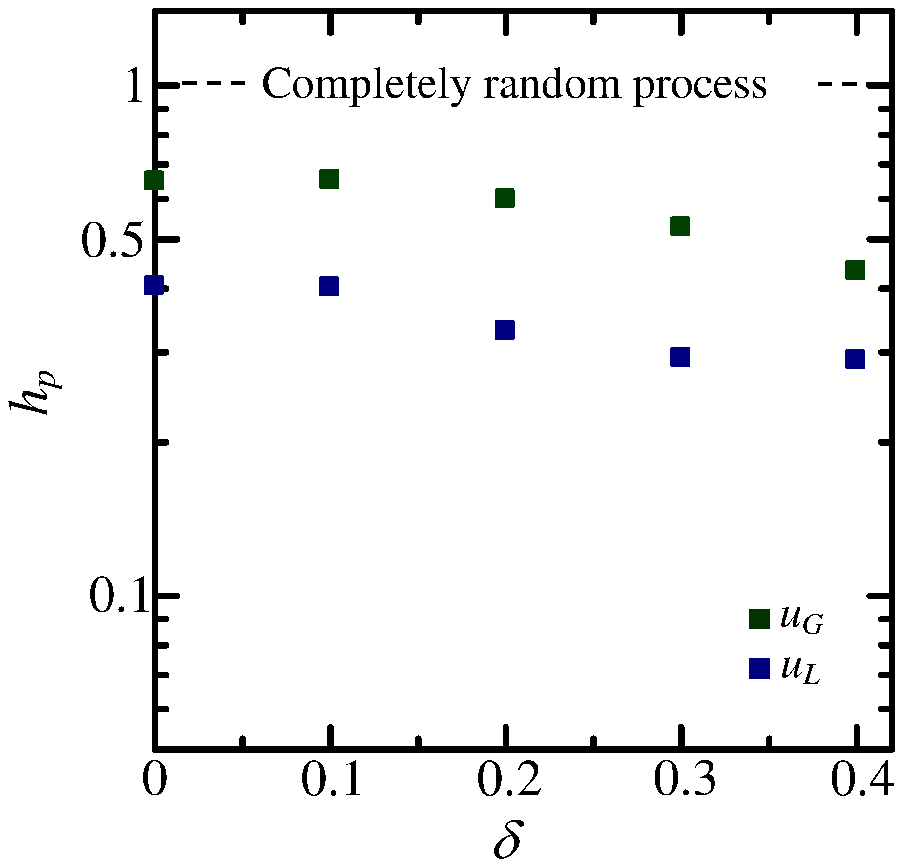}
\caption{Variations of $h_p$ for $u_L$ and $u_G$
computed by using different values of  $\delta$.}
\label{fig:19 permutation}
\end{figure}

\section{Conclusions}
\label{Sec:conclusions} We have presented a detailed and systematic study of
the dynamics of the spatiotemporal behavior of the gKS equation by making use
of statistical methods and time-series analysis based on chaos theory,
including a nonlinear forecasting method. The gKS equation is a
generalization of the KS equation to include dispersion, and has been used as
a prototype for the study of pattern formation dynamics and spatiotemporal
complexity in active dispersive-dissipative media. The spatiotemporal
solution of the gKS equation exhibits a rich dynamics that strongly depends
on the parameter $\delta$ controlling the strength of the dispersion. In
particular, one observes spatiotemporal chaos at low values of $\delta$, the
emergence of localized structures (pulses) constantly interacting with each
other at moderate values (regime also known as interfacial turbulence), and
periodic solutions at large values of $\delta$.

We have defined two different temporal magnitudes to be analyzed: a local
signal, the extracted solution in the middle of the system, and a global
measure, the second moment of the solution. This allows for two different
ways, a local and a global, of accessing the dynamical properties of the
system providing then a complete picture (usually in experiments one has
access to one of the two only). The local analysis has revealed that the
high-dimensional chaos regime ($\delta\lesssim 0.2$) is characterized by a
completely scattered graph in the phase space, a power-law decay PSD, and a
very short predictability time. As the dispersion parameter is increased, the
spatiotemporal solution starts to exhibit the emergence of solitary waves
(pulses) which continuously and chaotically interact with each other. This
corresponds to a low-dimensional chaos regime exhibiting a structured graph
with a strange attractor in the phase space, and with a typical critical
predictability time up to which the solution can be predicted. For larger
values of $\delta$ the critical predictability time starts to increase as the
solution approaches a periodic steady state ($\delta>1.8$).

We observe similar behavior for the global analysis but the transitions from
high-dimensional to low-dimensional chaos and finally to periodic solutions
occur at larger values of $\delta$, namely at $\delta\sim 1$ and
$\delta\sim 2.0$, respectively. This is a consequence of the global spatial
average we consider since any sign of localized structure is lost during such
averaging procedure. In this sense we can conclude that the global measure is
more chaotic than the local signal. This is also confirmed by looking at
the permutation entropy of the temporal signals observing that the global
signal has larger entropy than the local one.

We have also contrasted the results from the deterministic gKS equation to a
pure stochastic process (fractional Brownian), observing that one needs to
look at the increment process of the signal in order to distinguish between
stochastic and high-dimensional chaotic signals. In particular, we have
observed that while both the gKS solution at low values of $\delta$ and the
fractional Brownian motion exhibit a very short predictability time for the
original signal, the increment process shows zero predictability properties
for the stochastic process and short predictability properties for the
chaotic signal. To the best of our knowledge this is the first study to
provide a comprehensive interpretation of the dynamical state of the gKS
equation from the viewpoint of nonlinear forecasting.

Finally, there is a number of interesting questions related to the analysis
presented here. For instance, how can one use elements from stochastic
processes (e.g.~in \cite{Sebastian2013}) to appropriately modify the
nonlinear forecasting method when the gKS equation is postulated for a
nonlinear process for which the precise underlying model is not available.
Another interesting problem would be the extension of the methodology
proposed here to more involved equations, such as those describing the
dynamics of the falling film away from criticality. We shall examine these
and related issues in future studies.

\nonumsection{Acknowledgments} \noindent
We thank Dr. G. A. Pavliotis and Prof. D. T. Papageorgiou for useful comments
and suggestions. HG thanks the Department of Chemical Engineering of Imperial
College for hospitality during a sabbatical visit. He was partially supported
by a ``Grant-in-Aid for Young Scientists (A)" from the Ministry of Education,
Culture, Sports, Science and Technology of Japan (MEXT)" and from the JGC-S Scholarship Foundation. 
We acknowledge financial support from European Research Council (ERC) Advanced Grant No.
247031.


\end{document}